\documentclass[12pt]{iopart}

\usepackage{iopams,dsfont,bm}  
\usepackage{graphicx,color}
\usepackage{epsfig,amssymb}
\usepackage{amsthm}

\newcommand{\Comment}[1]{}
\newcommand{\Eq}[1]{Eq.~(\ref{#1})}

\theoremstyle{definition}
\newtheorem{definition}{Definition}[section]

\newcommand{\removemodulus}[1]{}

\newcommand{\usat}{\texttt{UNIQUE-SAT}}

 \newcommand{\ket}[1]{|{#1}\rangle}
 \newcommand{\bra}[1]{\langle{#1}|}
 \newcommand{\braket}[2]{\langle{#1}|{#2}\rangle}
\newcommand{\ip}[2]{\langle #1 ~|~ #2 \rangle}

\newcommand{\C}{\mathbb{C}}
\newcommand{\R}{\mathbb{R}}

\newcommand{\fnorm}[1]{\mathsf{N}(#1)}

\newcommand{\Fp}{\mathbf{F}_{p}}

\newcommand{\Fpp}{\mathbf{F}_{p^2}}

\newcommand{\Fq}{\mathbf{F}_{q}}

\newcommand{\Fpx}[1]{\mathbf{F}_{#1}}
\newcommand{\Fppx}[1]{\mathbf{F}_{{#1}^2}}

\makeatletter
\def\imod#1{\allowbreak\mkern8mu({\operator@font mod}\,\,#1)}
\makeatother

\newcommand{\upr}[1]{\uppercase\expandafter{#1}}

\def\fh{\mathfrak{h}}

\newcommand{\Hat}[1]{\hat{\bf #1}}
\newcommand{\Vec}[1]{{\bf #1}}

\newcommand{\VVec}[1]{\vec{\bf #1}}

\newcommand{\mathReal}{\mathbb{R}}
\newcommand{\mathComplex}{\mathbb{C}}

\newcommand{\Sphere}[1]{\mathbf{S}^{#1}}
\newcommand{\CP}[1]{\mathbf{CP}^{#1}}  
\newcommand{\DCP}[1]{\mathbf{DCP}^{#1}} 

\newcommand{\gmult}{*}

\begin{document}

\title{Geometry of Discrete Quantum Computing}

\author{Andrew J. Hanson}
\address{School of Informatics and Computing, Indiana University, Bloomington,
IN 47405, USA}
\author{Gerardo Ortiz}
\address{Department of Physics, Indiana University, Bloomington,
IN 47405, USA}
\author{Amr Sabry}
\address{School of Informatics and Computing, Indiana University, Bloomington,
IN 47405, USA}
\author{Yu-Tsung Tai}
\address{Department of Mathematics, Indiana University, Bloomington,
IN 47405, USA}



\begin{abstract}
  Conventional quantum computing entails a geometry based on the description
  of an $n$-qubit state using $2^{n}$ infinite precision complex numbers
  denoting a vector in a Hilbert space.  Such numbers are in general
  uncomputable using any real-world resources, and, if we have the idea of
  physical law as some kind of computational algorithm of the universe, we
  would be compelled to alter our descriptions of physics to be consistent
  with computable numbers.  Our purpose here is to examine the geometric implications
  of using finite fields $\Fp$ and finite complexified fields $\Fpp$ (based
  on primes $p$ congruent to $3 \imod{4}$) as the basis for computations in a
  theory of discrete quantum computing, which would therefore become a
  computable theory.  Because the states of a discrete $n$-qubit system are
  in principle enumerable, we are able to determine the proportions of entangled
  and unentangled states.  In particular, we extend the Hopf fibration that
  defines the irreducible state space of conventional continuous $n$-qubit
  theories (which is the complex projective space $\CP{2^{n}-1}$) to an
  analogous discrete geometry in which the Hopf circle for any $n$ is found
  to be a discrete set of $p+1$ points.  The tally of unit-length $n$-qubit
  states is given, and reduced via the generalized Hopf fibration to $\DCP{2^{n}-1}$, 
  the discrete analog of the complex projective space,  which
  has  $p^{2^{n}-1} (p-1)\,\prod_{k=1}^{n-1} ( p^{2^{k}}+1)$  %
  irreducible states.  Using a measure
  of entanglement, the {\it purity\/}, we explore the entanglement features
  of discrete quantum states and find that the $n$-qubit states based on the
  complexified field $\Fpp$ have $p^{n} (p-1)^{n}$ unentangled states (the
  product of the tally for a single qubit) with purity 1, and they have
  $p^{n+1}(p-1)(p+1)^{n-1}$ maximally entangled states with purity zero.
\end{abstract}


\maketitle

\section{Introduction}

\medskip {\it Conventional quantum computing\/} (CQC) is appealing
because it expands our horizons on the concepts of computing in
general.  The fundamental principles of CQC broadly influence computer
science, physics, mathematics, and logic.  Not only would Turing have
been fascinated by the implications of quantum computing for his own
theory of computation, but he would also have been intrigued by the
apparent absence of any further possible extensions.  In this note we
go one step further, and study the beginnings of a fundamental
consistent framework for {\it discrete quantum computing\/} (DQC).
Our basic results in this paper include a detailed construction and
analysis of the irreducible $n$-qubit states in DQC, a novel
analysis of the structure of the discrete generalized Bloch sphere for
$n$-qubits and a study of entanglement in the discrete domain.  

Research on theoretical quantum computing focuses on two distinct
aspects, algorithms and geometry.  Since quantum computing contains
features and components quite different from classical computational
methods, the exploration of algorithms, computation, and the theory of
computational methods is essential, and includes the study of topics
such as the Deutsch-Jozsa algorithm whose task is to determine if a
function is constant or balanced with preternatural speed, and
Grover's algorithm for searching a database with the square root of
the number of queries needed classically.  But another essential
branch of quantum computing research is the investigation of the
nature of {\it states themselves\/}, the geometry of the spaces
describing the $n$-qubit states upon which algorithms eventually act;
the properties of such spaces are important in their own right, long
before they are used in algorithms.  Understanding these properties
serves, for example, to explicate the nature of irreducible states
(when all wave-function symmetries are eliminated), and exposes the
nature of entangled states, a phenomenon completely absent from any
non-quantum geometrical framework.  The geometric aspects of
conventional quantum theory and quantum computing are the subject of a
vast literature, and entire books (see, e.g., \cite{GeomQStates2008})
have been devoted to quantum geometry and its relation to
entanglement.  An extensive picture of the geometry of conventional
quantum computing has emerged, showing that the complex projective
spaces $\CP{2^n - 1}$ precisely embody the irreducible states of an
$n$-qubit quantum circuit element, and, in addition, permit the
explicit study of the actual paths in the irreducible state space that
correspond to idealized quantum operations.

Our contribution, which involves issues possibly less familiar to
readers of the algorithm-centered literature, is to extend the path of
the corpus of ``conventional'' geometry-based quantum computing
research into the discrete domain.  We start with a finite
complexified Galois field $\Fpp$ replacing the complex fields used in
the existing literature for the geometry of quantum computing
(e.g., \cite{GeomQStates2008, mosseri2001}) and examine the implications of
calculating the geometric properties of $n$-qubit states with
coefficients defined in discrete Galois  fields.  Our work for the first time
explicates a rigorous approach to $n$-discrete-qubit complex geometry
and the resulting discretized complex projective spaces.  We rederive
some of the basic results of discrete complex
mathematics introduced by Vladimir Arnold \cite{Arnold}, and extend
these to a discrete attack on the entire spectrum of geometric
problems appearing in the conventional quantum computing literature.
Among the new insights that appear in our approach are explicit
relative measures for counting the numbers of unentangled, partially
entangled, and maximally entangled states, along with the dependence
of these measures on the size of the chosen discrete fields.  All of
this structure is concealed by the infinite precision of real numbers
in conventional quantum computing, and thus the discrete methods
provide ways of understanding the resources of quantum computing and
isolating the relations between resources and problem size that cannot
be studied in any other fashion.  These are significant new results,
whose ultimate implications cannot be trivially predicted.

This work is given impetus by the fact that the great majority of the laws
of physics are formulated as equalities (more appropriately, as
isomorphisms) between different physical observables. For instance,
Newton's second law of classical mechanics equates the force acting on
a system to its rate of change of momentum.  Another type of law is
the second law of thermodynamics, which asserts that the entropy of a
system increases as the system evolves in time, with a corresponding
mathematical formulation in terms of an inequality.  It is certainly
appealing to relate the laws of physics described in this way to
computational algorithms.  However, an important observation is that
the laws of physics are in general implicitly formulated in terms of
uncomputable numbers.  We therefore concern ourselves with the issue
of whether conventional quantum mechanics is physical, or whether
perhaps extremely large discrete quantum theories that contain only
computable numbers are at the heart of our physical universe.
Imagining that physical laws might ultimately require computable
numbers provides a compelling motivation for the research program in
DQC to which this paper is devoted.

Of specific relevance to our topic is the fact that the title of
Turing's seminal 1937 paper \cite{Turing37} was ``On Computable
Numbers\ldots .'' The idea of computable numbers is of foundational
significance in computer science and has had a significant impact on logic.
However, despite arguments and challenges noted by prominent researchers
\cite{Minsky67,Chaitin,Feynman}, most mathematical
models depend completely on uncomputable numbers, that is, the
continuum of real (or complex) numbers; the mathematical framework of
conventional quantum mechanics is based on Hilbert spaces, which have
uncomputable numbers as their underlying field.
In the words of Rolf Landauer~\cite{Landauer}, 
\begin{quote}
\ldots the real world is unlikely to supply us with
  unlimited memory of unlimited Turing machine tapes.  Therefore,
  continuum mathematics is not executable, and physical laws which
  invoke that can not really be satisfactory \ldots
\end{quote}
  Here we explore a
further plausible principle of quantum computing --- the hypothesis that,
because of the finiteness of resources in the universe, the domain of
physical computation (thus including quantum mechanics) could be
restricted to computable numbers and finite fields.

When we began this research program some years ago, our starting point,
like that of Schumacher and Westmoreland~\cite{SchuWest2010}, was to
investigate the properties of a version of quantum mechanics obtained by
instantiating the mathematical framework of Hilbert spaces with the smallest
finite field of booleans instead of the field of complex numbers. That ``toy
model'' was called \emph{modal quantum mechanics} by Schumacher and
Westmoreland. Our first result~\cite{James2011} was to explicate the
associated model of computing as a conventional classical model of relational
programming with one twist that is responsible for all the ``quantum-ness.''
More precisely, we isolated the ``quantum-ness'' in the model in one
operation: that of merging sets of answers computed by several
alternative choices in the relational program. In the classical world, the
answers are merged using a plain union; in modal quantum computing, the
answers are merged using the \emph{exclusive union}, which is responsible
for creating quantum-like interference effects.

Despite the initial expectations that modal quantum computing would be a
``toy'' version of CQC, we showed --- in a surprising development --- that
modal quantum computing exhibited \emph{supernatural} computational
power. More precisely, we showed that the \usat\ problem (the question of
deciding whether a given boolean formula has a satisfying assignment,
assuming that it has at most one such assignment) can be solved
\emph{deterministically} and in a \emph{constant number of black box
  evaluations} in modal quantum computing. We traced this supernatural power
to the fact that general finite fields lack the geometrical structure
necessary to define unitary transformations, and proposed instead the
framework of \emph{discrete quantum theory}~\cite{HOSW2011}. This framework
is based on complexified Galois fields (see, for example,
Arnold~\cite{Arnold}) with characteristic $p=4\ell+3$ for $\ell$ a non-negative
integer (i.e., $p \equiv 3 \imod{4}$), which recover enough geometric
structure to define orthogonality and hence allow the definition of Hermitian
dot products and unitary transformations.

Discrete quantum theories eliminate the particular supernatural algorithm for
\usat. They however still allow subtle supernatural algorithms that
depend on the precise relation of the characteristic of the field $p$ and the
number of qubits used in the calculation. In particular, we were able to show
that supernatural behavior can happen in versions of \usat\ for a database of
size $N$ if the characteristic $p$ of the field divides $(2^N -
1)$~\cite{HOSW2011}. 

This paper
explores the notions above in detail from first principles. 
We will focus our attention on the specific challenge that
confronts any attempt to build an $n$-qubit quantum computing structure based
on the classical mathematical domain of {\it finite fields\/}, and
particularly on the shift in the concepts of geometry as one transitions from
the continuous case (CQC) to the discrete case (DQC).  The fundamental
mathematical structure that we shall refer to throughout is the finite field
$\Fpx{p^{r}}$, where $p$ is a prime number, with some possible restrictions,
and $r\ge 1$ is an integer. We shall see below that $\Fpp$ in particular will
give us a precise discrete analog to the continuous complex probability
amplitude coefficients of conventional $n$-qubit quantum states.

Our task is then to extract some minimal subset of
the familiar geometric properties of CQC in the context of the
unfamiliar geometric properties of DQC.  It does not take long to
discover a litany of issues such as the following:

\begin{itemize} 
\item {\bf CQC is based on continuous (typically uncomputable) complex state
    coefficients\/} in the complex number field $\C$, whose absolute squares
  are continuous (typically uncomputable) real probabilities in $\R$ that are
  {\it ordered\/}: one can always answer the question asking whether one
  probability is greater than another.  In DQC, we still have (a discrete
  version of) complex numbers in $\Fpp$, and their absolute squares still
  have real values in $\Fp$; however, in $\Fp$, there is no transitive {\it
    order\/} --- all real values repeat modulo $p$, and, without additional
  structure, we cannot, {\it even in principle\/}, tell what the ordering
  should be (e.g., for $p=3$, the label set $\{-1,0,1\}$ is just as good as
  $\{0,1,2\}$).  There are ways to label ``positives'' and ``negatives'' in
  the finite field $\Fp$, and ways to assign ordered local neighborhoods
  under certain restrictive conditions, but we still have no consistent way
  to order the numbers in an entire field.

\item {\bf In CQC there is no distinction between geometric proximity
    of vectors and probability of closeness.}  The calculation for the
  two concepts is the same.  In DQC, there is no notion of closeness
  of vectors that can be computed by inner products or probabilities,
  although there are deep geometric structures on discrete lattices.  One
  of our challenges is therefore to tease out some meaning from this
  geometry despite its failure to support the expected properties of
  such common operations as inner products that are compatible with
  our intuitions from real continuous geometry.

\item {\bf In ordinary real and complex geometry, we have continuous
  notions of trigonometry.\/}  Additional notions implying continuous
geometry for ordinary number fields include linear equations whose
solutions are continuous lines, quadratic equations whose solutions
are manifolds such as spheres, and continuous-valued measurable
quantities such as lengths of line segments, areas of triangles,
volumes of tetrahedra, etc.  In a discrete real or complex lattice
corresponding to $\Fp$ or $\Fpp$, analogs of many of these familiar
geometric structures exist, but they have unintuitive and unfamiliar
properties. We will expand on these geometric structures in a future 
publication.
\end{itemize}

We proceed in our exposition first by reviewing the underlying geometry of
continuous $n$-qubit states in CQC, including a discussion of the properties
of entanglement.  Our next step is to review the often non-trivial technology
of real and complex discrete finite fields.   Finally, we examine the features of discrete state geometry for
$n$-qubits, including entanglement, as they appear in the context of states
with discrete complex ``probability amplitude'' coefficients.
In particular, we extend  the Hopf fibration of CQC 
(which is the complex projective space $\CP{2^{n}-1}$) to a
discrete geometry in which the Hopf circle contains 
$p+1$ points.  The resulting discrete complex projective space $\DCP{2^{n}-1}$ 
has $p^{2^{n}-1} (p-1)\,\prod_{k=1}^{n-1} ( p^{2^{k}}+1)$
irreducible states, $p^{n} (p-1)^{n}$ 
of which are  unentangled and $p^{n+1}(p-1)(p+1)^{n-1}$ maximally entangled states.

\Comment{  
\begin{quote}
{\it \footnotesize {\normalsize Sabry: further notes:}
[Then we introduce QC and why Turing would be interested in QC along the
lines of the current introduction.]\\[.1in]
[Then we argue like the current intro that it turns out that the real
numbers are hiding geometry, probability, etc. and we are teasing the
geometry piece...] }
\end{quote}
\begin{quote}
{\it \footnotesize {\normalsize Ortiz: Further motivation:}
[Motivate Physical law and its connection to algorithms. 
Kinematics  $-->$ Data structure , 
Dynamics $-->$ Algorithms ]\\[.1in]
[Use also discrete Grover algorithm in the motivation] }
\end{quote}
}  

\section{Continuous Quantum Geometry}
\label{CQC1qubitBloch.sec}

\bigskip
Conventional quantum computation is  described by the following:
\begin{enumerate}
\item[(i)] $D=2^{n}$ orthonormal basis vectors of an $n$-qubit state,
\item[(ii)]  the  normalized $D$  complex probability amplitude 
coefficients describing the contribution of each basis vector,
\item[(iii)]   a set of probability-conserving unitary matrix
  operators that suffice to describe all required state
  transformations of a  quantum circuit,
\item[(iv)]   and a measurement framework.  
\end{enumerate}
We remark that there are many things that are assumed in
CQC, such as the absence of zero norm states for non-zero vectors,
and the decomposition of complex amplitudes into a pair of
ordinary real numbers.  One also typically assumes the existence of a
Hilbert space with an orthonormal basis, allowing us to write 
$n$-qubit {\it pure\/} states
in general as Hilbert space vectors with an Hermitian inner product:
\begin{eqnarray} 
 \ket{\Psi} & =&  \sum_{i=0}^{D-1} \alpha_{i}  \ket{i} \ .
\end{eqnarray}
Here  $\alpha_{i} \in \mathComplex$ are complex probability amplitudes,  $\VVec{\alpha} \in
\mathComplex^{D}$, and the $\{\ket{i}\}$ is an orthonormal basis of
states obeying
\begin{eqnarray}
\braket{i}{k}=\delta_{ik}  \ .
\end{eqnarray}

The meaning of this is that any  state $\ket{\Phi}=\sum_{i=0}^{D-1} 
\beta_{i}  \ket{i}$ can be projected  onto
another state $\ket{\Psi}$ by writing
\begin{eqnarray} 
 \braket{\Phi}{\Psi} & = & \sum_{i=0}^{D-1} \beta^{*}_{i} \alpha^{\;}_{i} \ ,  
\end{eqnarray}
thus quantifying the proximity of the two states.  (Here ${}^{*}$
denotes complex conjugation.)  This is one of many
properties we take for granted in continuum quantum mechanics that
challenge us in defining a discrete quantum geometry.

In this paper, we focus on the discrete geometric issues raised by the
properties (i) and (ii) given above for CQC, and leave for another 
time the important issues of (iii),
(iv), and such conundrums as probabilities, zero norms,  and dynamics
in the theory of DQC.


To facilitate the transition to DQC carried out in later sections, we
concern ourselves first with the properties of the simplest possible
abstract state object in CQC, the single qubit state.

\subsection{The single qubit problem}
A single qubit already provides access to a wealth of geometric
information and context.  We write the  single qubit state as
\begin{equation}
\begin{array}{lr}\ket{\psi_{1}} =\alpha_{0} \ket{0} + \alpha_{1} \ket{1} \ &
  \   \alpha_{0}, \alpha_{1} \in  \C \end{array} \ .
\label{A1.eq}
\end{equation}

A convenience for probability calculations and a necessity for
computing relative state properties is the normalization condition
\begin{equation}
\|\psi_{1} \|^{2}= | \alpha_{0}|^{2} +| \alpha_{1}|^{2} = 
\alpha_{0}^* \alpha_{0}+ \alpha_{1}^* \alpha_{1}=1 \ ,
\label{A2.eq}
\end{equation}
which identifies $\alpha_{0}$ and $\alpha_{1}$ as (complex)
probability amplitudes and implies the conservation of probability in
the closed world spanned by $\{\ket{0}, \ket{1}\}$.  Note that we
distinguish for future use the {\it norm\/} $\| \cdot\|$ of a vector
from the {\it modulus\/} $|\cdot |$ of a complex number.  Continuing,
we see that if we want only the irreducible state descriptions, we
must supplement the process of computing \Eq{A2.eq} by finding a way
to remove the distinction between states that differ only by an
overall phase transformation $e^{i \phi}$, that is,
\begin{equation} 
( \alpha_{0}, \alpha_{1}) \sim (e^{i \phi}\alpha_{0}, e^{i
  \phi}\alpha_{1}) \ .
\label{A3.eq}
\end{equation}
This can be accomplished by the Hopf fibration, which we can write
down as follows: let
\begin{equation} 
\begin{array}{cc} \alpha_{0} = x_0 + i y_0 ,&  \alpha_{1} = x_1 + i y_1 
\end{array} \ .
\label{A4.eq}
\end{equation}
Then \Eq{A2.eq} becomes the condition that the four real
variables describing a qubit denote a point on the three-sphere
$\Sphere{3}$ (a 3-manifold) embedded in $\R^{4}$:
\begin{equation} 
x_{0}^{\ 2} +  y_{0}^{\ 2} + x_{1}^{\ 2} + y_{1}^{\ 2} = 1 \ .
\label{A5.eq}
\end{equation}

\begin{figure}[htb]
\centerline{
\includegraphics[width=0.48\columnwidth]{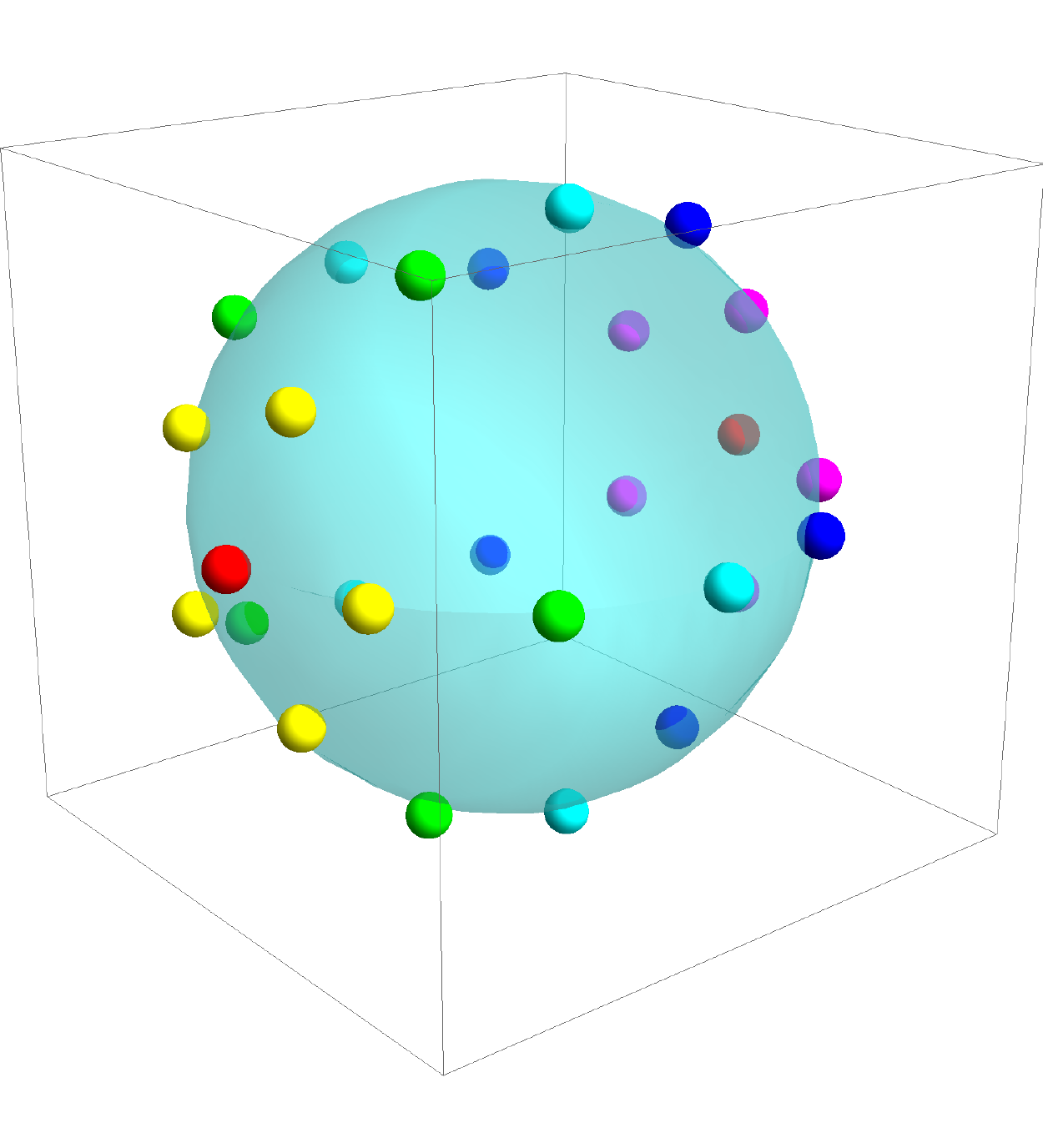}\hspace{0.03in}
\includegraphics[width=0.48\columnwidth]{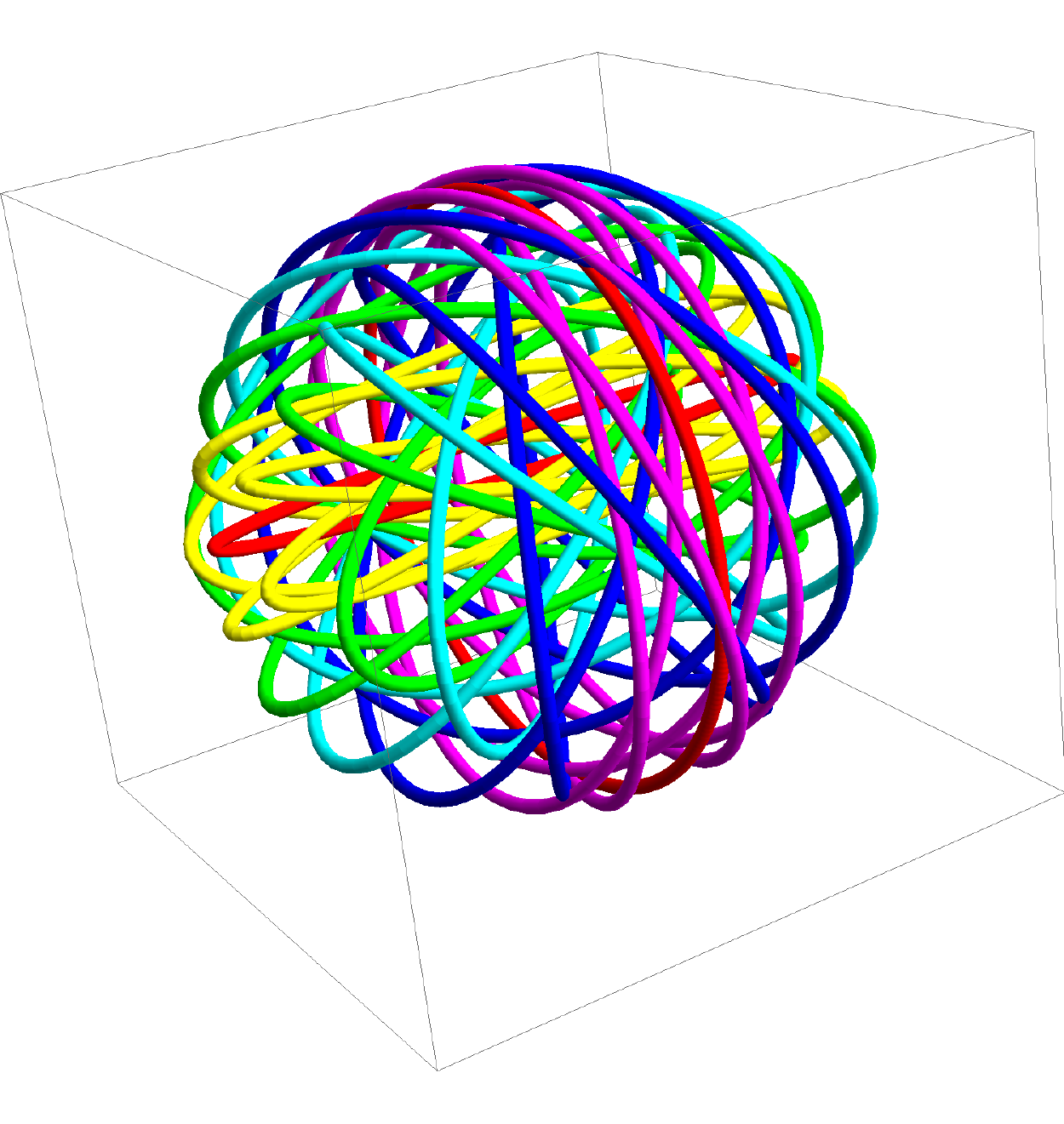}}
\centerline{\hspace{1in} (a) \hfill (b) \hspace{1in}}
 \caption{(a) The sphere represented by \Eq{A7.eq}, which is
the irreducible space of one-qubit states, along with a representative
set of points on the sphere.
(b) Representation of the Hopf fibration as a family of circles
(the paths of $e^{i \phi}$), each corresponding to a single point
on the sphere in (a).  Points in (a) are color coded corresponding to
circles in (b), e.g., one pole contains the red elliptical
circle that would become an infinite-radius circle in a slightly different
projection,  and the opposite pole corresponds to the large perfectly
round red circle at the equator.}
 \label{hopfFibers.fig}
\end{figure}

There is a family of 6 equivalence classes of quadratic maps that take
the remaining 3 degrees of freedom in \Eq{A5.eq} and reduce them
to 2 degrees of freedom by effectively removing $e^{i\phi}$
(``fibering out by the circle $\Sphere{1}$'').  The standard form of
this class of maps (``the Hopf fibration'') is
\begin{eqnarray} 
 X & = & 2\, \mathrm{Re}\   \alpha_{0} \alpha_{1}^{*} = 2 x_0 x_1 + 2
 y_0 y_1 \nonumber \\
 Y & = & 2\, \mathrm{Im}\   \alpha_{0} \alpha_{1}^{*} = 2 x_1 y_0 - 2 x_0 y_1 \\
 Z & = &   | \alpha_{0}|^{2} - | \alpha_{1}|^{2} =  x_{0}^{\ 2} +
   y_{0}^{\ 2} - x_{1}^{\ 2} - y_{1}^{\ 2}  \ .  \nonumber
\label{A6.eq}
\end{eqnarray}
These transformed coordinates obey
\begin{equation} 
\|\Vec{X}\|^{2} = X^2 + Y^2 + Z^2 = \left(| \alpha_{0}|^{2} +|
  \alpha_{1}|^{2}\right)^{2} = 1  
\label{A7.eq}
\end{equation}
and therefore have only two remaining degrees of freedom describing
all possible distinct one-qubit quantum states.  In Figure
\ref{hopfFibers.fig} we illustrate schematically the family of circles
{\it each one of which is collapsed to a point\/} $(\theta,\phi)$ on
the surface $ X^2 + Y^2 + Z^2=1$ by the Hopf map.

\begin{figure}[thb]
\begin{center}
\includegraphics[width=0.47\columnwidth]{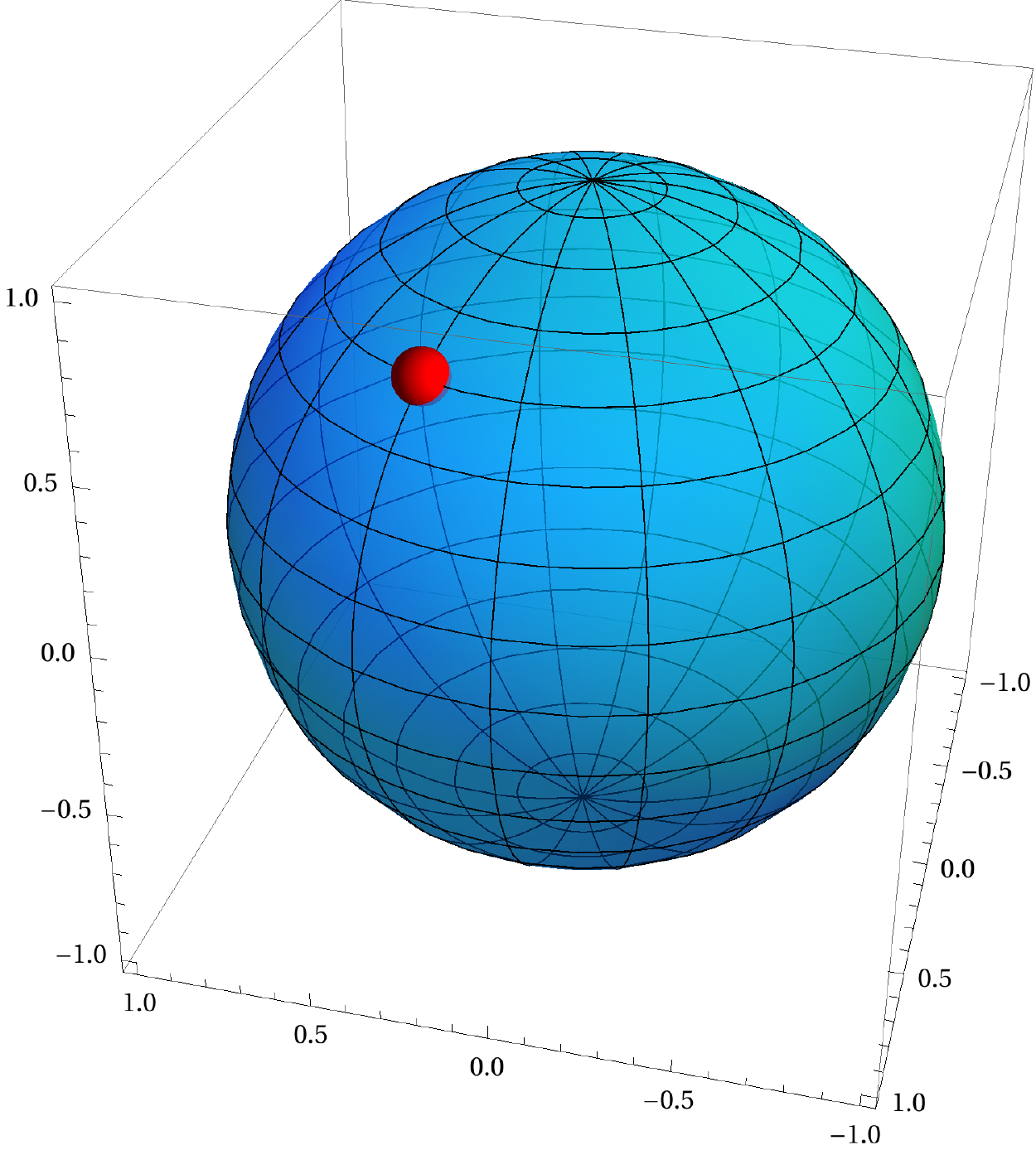}
\hspace{0.01in}
\includegraphics[width=0.47\columnwidth]{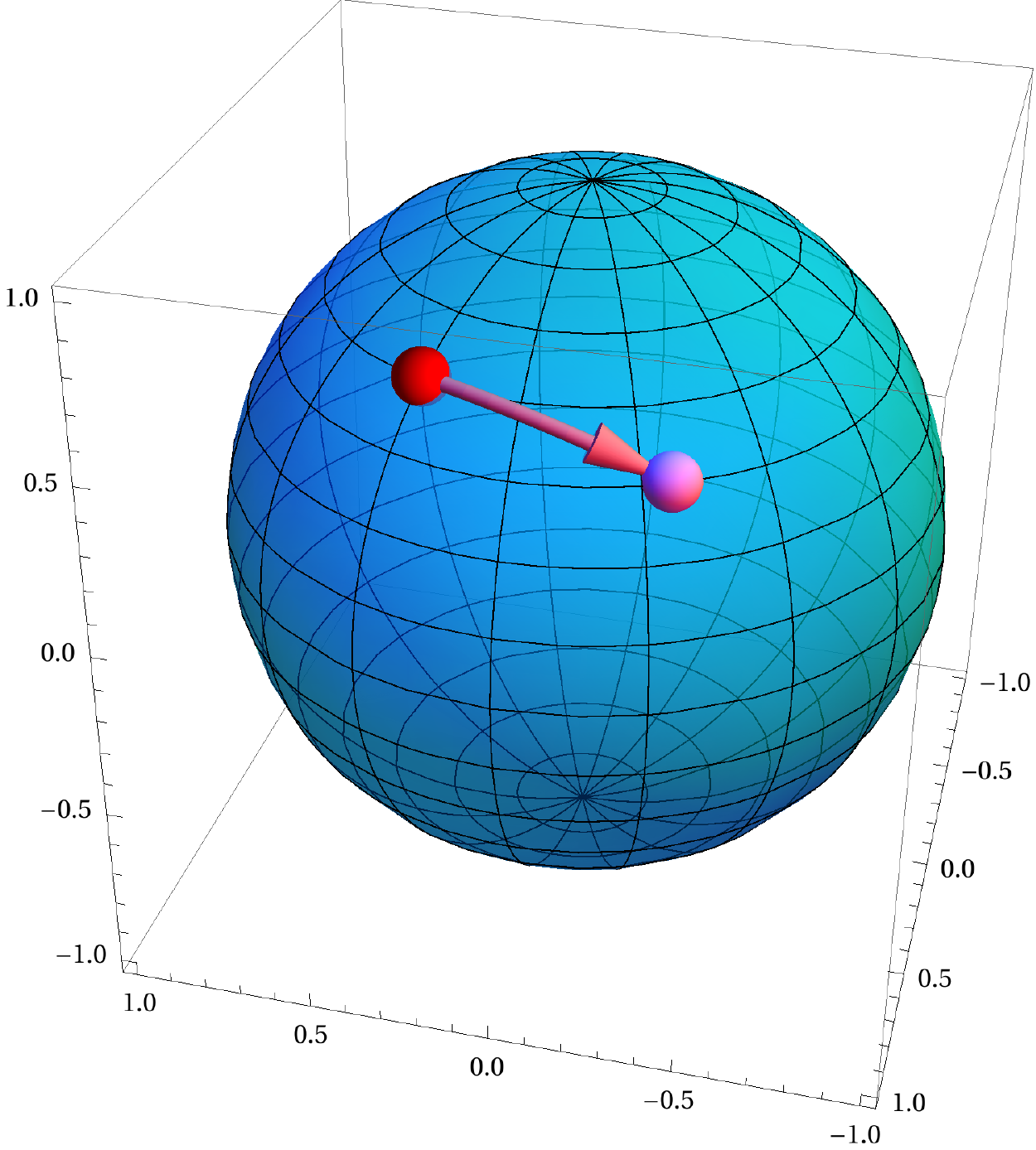}
\end{center}
\centerline{\hspace{1in} (a) \hfill (b) \hspace{1in}}
 \caption{(a) The conventional Bloch sphere with a unique state
represented by the point at the red sphere. (b) The geodesic
shortest-distance arc connecting two one-qubit quantum states.}
 \label{blochSphere.fig}
\end{figure}

  The resulting
manifold is the two-sphere $\Sphere{2}$ (a 2-manifold) embedded in
$\R^{3}$.  If we choose one of many possible coordinate systems
describing $\Sphere{3}$ via \Eq{A5.eq} such as
\begin{eqnarray} 
\hspace*{-0.5cm}
\left(x_{0}, y_{0},  x_{1},  y_{1}\right) &=& 
\left(\cos\frac{\theta +\phi}{2} \cos\frac{\psi}{2},\,  
\sin\frac{\theta +\phi}{2} \cos\frac{\psi}{2},\ \right . \nonumber \\
&&\left . 
\cos\frac{\theta -\phi}{2} \sin\frac{\psi}{2},\
\sin\frac{\theta -\phi}{2} \sin\frac{\psi}{2}\right) , 
\label{A8a.eq}
\end{eqnarray}
where 
   $0\leq \psi \leq \pi$, 
with  $0\leq \frac{\theta +\phi}{2} <  2 \pi$  and
$0\leq \frac{\theta -\phi}{2} < 2 \pi$,
we see that
\begin{eqnarray} 
\left(X,Y,Z \right) =
\left(\cos\phi \sin\psi,\,  \sin\phi \sin\psi,\,\cos\psi \right) \ .
\label{A8b.eq}
\end{eqnarray}
Thus the one-qubit state is independent of $\theta$, and we can choose
$\theta=\phi$ without loss of generality, reducing the form of the
unique one-qubit states to
\begin{equation} 
\ket{\psi_{1}} =  e^{i \phi} \cos\frac{\psi}{2} \ket{0} +
  \sin\frac{\psi}{2}\ket{1}  \ ,
\label{A9.eq}
\end{equation}
and an irreducible state can be represented as a point on a sphere,
as shown in Figure \ref{blochSphere.fig}(a).

Thus the geometry of a single qubit reduces to transformations among
points on $\Sphere{2}$, which can be parametrized in an infinite
one-parameter family of transformations, one of which is the geodesic
or minimal-length transformation.  Explicitly, given two one-qubit states
denoted by points $\Vec{a}$ and $\Vec{b}$ on  $\Sphere{2}$,  the
shortest rotation carrying the unit normal $\Hat{a}$ to the unit normal 
$\Hat{b}$ is the SLERP (spherical linear interpolation)
\begin{equation} 
S(\Hat{a},\Hat{b},t) = \Hat{a}\,\frac{\sin((1-t)\theta)}{\sin \theta}\, + \, 
\Hat{b}\,\frac{\sin(t \theta)}{\sin \theta}  \ ,
\label{slerp.eq}
\end{equation}
where $\Hat{a}\cdot \Hat{b} = \cos \theta$. 
Figure \ref{blochSphere.fig}(b) illustrates the path traced by a SLERP
between two irreducible one-qubit states on the Bloch sphere.  Because states
in CQC are defined by infinite precision real numbers, it is not possible,
even in principle, to make an exact state transition as implied by Figure
\ref{blochSphere.fig}(b).  In practice, one has to be content with
approximate, typically exponentially expensive, transitions from state to state.


\subsection{The $n$-qubit problem}
For $n$ qubits, the irreducible states are encoded in a similar family of
geometric structures known technically as the complex projective space
$\CP{D-1}$.  We obtain these structures starting with the $D = 2^{n}$
initially unnormalized complex coefficients of the $n$-qubit state basis
\begin{eqnarray}
\ket{\Psi}  &=& \sum_{i=0}^{D-1} \alpha_{i} \ket{i}  \ .
\label{N1.eq}
\end{eqnarray}
We then follow the analog of the one-qubit
procedure:
Conservation of probability  requires that the
norm of the vector  $\VVec{\alpha}$ be normalized to unity:
\begin{eqnarray}
 \braket{\Psi}{\Psi} = \| \VVec{\alpha}\|^{2}& =& 
    \sum_{i=0}^{D-1} |\alpha_{i}|^{2}\, =\, 1\ .
\label{probEq1.eq}
\end{eqnarray}
Thus the initial equation for the geometry of a quantum state describes
a {\it topological sphere\/} $\Sphere{2D-1}$ embedded in
$\mathReal^{2D}$.  To see this, remember that we can write the real
and imaginary parts of $\alpha_{i}$ as $\alpha_{i} = x_{i} + i y_{i}$,
so
\begin{eqnarray}
 \sum_{i=0}^{D-1} |\alpha_{i}|^{2} = \sum_{i=0}^{D-1}\left(
x_{i}^{\ 2} + y_{i}^{\ 2}\right) = 1
\end{eqnarray}
describes the locus of a $2D$-dimensional real unit vector in
$\mathReal^{2D}$, which is by definition $\Sphere{2D-1}$,
the $(2D -1)$-sphere, with $D=2^{n}$ for an $n$-qubit state.

This $\Sphere{2D-1}$ in turn is ambiguous up to the usual overall
phase, inducing an $\Sphere{1}$ symmetry action, and identifying $\Sphere{2D-1}$
as an $\Sphere{1}$ bundle, whose base space is the
$(D-1)$-complex-dimensional projective space $\CP{D-1}$.  There are
thus $2D-2$ irreducible real degrees of freedom ($D-1$ complex degrees
of freedom) for a quantum state with a
$D$-dimensional basis, $\{\ket{i}\, |\, i = 0,\ldots,D-1\}$.

In summary, the full space of a $D=2^{n}$-dimensional $n$-qubit quantum
state, including its overall phase defining its relationship to other quantum
states, is the topological space $\Sphere{2D-1}$.  For an isolated system,
the overall phase is not measurable, and eliminating the phase dependence in
turn corresponds to identifying $\Sphere{2D-1}$ as a circle bundle over the
base space $\CP{D-1}$, and therefore $\CP{D-1} = \CP{2^{n}-1}$ defines the
$2D-2$ intrinsic, irreducible, degrees of freedom of the isolated
$n$-qubit state's dynamics. In mathematical notation, this would be
written 

\[ \Sphere{1} \hookrightarrow \Sphere{2D-1}  \rightarrow \CP{D-1}\ ,\]
with  $D=2^{n}$ as usual.  For $n=1$, the single qubit, we have
$ 2^{n} - 1 = 2 -1 = 1$, and the base space of the circle bundle is
$\CP{1}=\Sphere{2}$, the usual Bloch sphere.  Note that only for
$n=1$ is this actually a sphere-like geometry due to an accident of
low-dimensional topology.

\subsection{Explicit $n$-qubit generalization of the Hopf fibration
  construction}   
For one qubit, we could easily solve the problem of reducing the full
unit-norm space to its irreducible components $\Vec{X}=(X,Y,Z)$
characterizing the Bloch sphere.  We have just argued that essentially
the same process is possible for $n$-qubits: in the abstract argument,
we simply identify the family of coefficients $\{\alpha_{i}\}$ as
being the same if they differ only by an overall phase $e^{i \phi}$.
However, in practice this is not a construction that is easy to
realize in a practical computation.  We now outline an explicit
algorithm for accomplishing the reduction to the irreducible $n$-qubit
state space $\CP{D-1}$; this construction will turn out to be useful
for the validation of our discrete results to follow below.

We begin by noting that a natural quantity characterizing an $n$-qubit
system is its {\it density matrix\/}, $\rho = \left[ \alpha_{i}
  \alpha^{*}_{j} \right]$, or
\begin{equation} 
\rho = \left[ \begin{array}{cccc}
| \alpha_{0}|^{2} &  \alpha_{0}  \alpha^{*}_{1} & 
        \cdots &  \alpha_{0}  \alpha^{*}_{D-1}\\
 \alpha_{1}  \alpha^{*}_{0} & | \alpha_{1}|^{2} &  
        \cdots &  \alpha_{1}  \alpha^{*}_{D-1}\\
 \vdots &  \vdots & \ddots & \vdots \\
  \alpha_{D-1}  \alpha^{*}_{0} & \cdots & 
    \alpha_{D-1}  \alpha^{*}_{D-2} & | \alpha_{D-1}|^{2}
\end{array} \right] \ .
\label{nqDensity.eq}
\end{equation}

We can now use the complex generalization of the classical Veronese
coordinate system for projective geometry to remove
the overall phase ambiguity $e^{i \phi}$ from the $n$-qubit
states.  If we take a particular weighting of the elements of the
density matrix $\rho$, we can construct a {\it unit vector\/} of real
dimension  $D^2$ with the form:
\begin{equation}
\Vec{X} = ( |\alpha_{i}|^{2},\ldots, \sqrt{2} \,  \mathrm {Re}\   \alpha_{i}
\alpha_{j}^{*}, \ldots, \sqrt{2} \, \mathrm{Im}\   \alpha_{i}
\alpha_{j}^{*},\ldots )\ , 
\label{nqCoords.eq}
\end{equation}
where
\begin{eqnarray}
 \Vec{X}\cdot\Vec{X} = \left( \sum_{i=0}^{D-1} \left |
    \alpha_{i}\right | ^{2}\right) ^{2} = 1 \ . 
\end{eqnarray}
This construction gives an explicit embedding of the $(D-1)$-dimensional
complex, or $(2 D -2)= (2^{n+1}-2)$-dimensional real, object in a real
space of dimension $D^2 = 2^{2 n}$.  However, this is somewhat subtle
because the vector is of unit length, so technically the embedding
space is a sphere of dimension $D^2-1 = 2^{2 n} -1$ embedded in
$\R^{D^2}$;  the one-qubit irreducible states could be represented in
a 4D embedding, but the magnitude of every coordinate would be
one; furthermore, the object embedded in the resulting $\Sphere{3}$ is
indeed $\Sphere{2}$ because we can fix one complex coordinate to be
unity, and let one vary, giving a total of two irreducible
dimensions.  In fact one must choose {\it two\/} coordinate patches,
one covering one pole of $\Sphere{2}$ with coordinates
\begin{eqnarray} \begin{array}{ccc}
\alpha_{0} &=& 1 + 0 i\\
\alpha_{1} &=& x_{1} +  i y_{1} \end{array} 
\end{eqnarray}
and the other patch covering the other pole of $\Sphere{2}$ with
coordinates
\begin{eqnarray}  \begin{array}{ccc}
\alpha_{0} &=& x_{0} + i y_{0}\\
\alpha_{1} &=& 1 + 0 i \ .\end{array} 
\end{eqnarray} 

 We finally see that the irreducible $n$-qubit
state space $\CP{D-1}$ is described by $D$ projectively equivalent
coordinates, one of which can always be scaled out to leave $(D-1)$
actual (complex) degrees of freedom.  We must choose, in turn, $D$ different
local sets of complex variables defined by
taking the value  $\alpha_{k}=1$, with $k=0,\ldots,D-1$, and allowing the
remaining $D-1$ complex (or $2 D -2$ real) variables to run free. 
No single set of coordinates will work, since the submanifold including
$\alpha_{k} = 0$ is undefined and another coordinate system must be
chosen to cover that coordinate patch.  This is a standard feature of
the topology of non-trivial manifolds such as $\CP{D-1}$ (see any
textbook on geometry \cite{Berger}).

\subsection{The geometry of entanglement}
Entanglement may be regarded as one
of the main characteristics distinguishing quantum from classical
mechanics. Entanglement  involves 
quantum correlations such that the measurement outcomes in one
subsystem are related to the measurement outcomes in another
one. Within the standard framework, given a quantum system
composed of $n$ qubit subsystems, a pure state of the
total system $\ket{\Psi}$ is said to be entangled if it cannot be
written as a product of states of each subsystem. That is, 
a state  $\ket{\Psi}$ is entangled if 
\begin{eqnarray}
\ket{\Psi}
\ne \ket{\psi_1}\otimes \cdots \otimes \ket{\psi_j} \otimes \cdots \otimes \ket{\psi_n},
\end{eqnarray}
where
$\ket{\psi_j}$ refers to an arbitrary state of the $j$-th qubit,  
and $\otimes$ represents the tensor product. 
This is equivalent to saying that if one
calculates the reduced density operator $\rho_j$ of the $j$th
subsystem by tracing out all the other subsystems,
$\rho_j= {\sf tr}_{\{1,\cdots, j-1,j+1,\cdots, n\}} (\rho)$, with
$j=1,\cdots, n$ and $\rho=\ket{\Psi}\bra{\Psi}$, the normalized state
$\ket{\Psi}$, $\langle \Psi | \Psi \rangle=1$, is entangled if and only if at least one
subsystem state is {\it mixed\/}; i.e., ${\sf tr}_j (\rho^2_j) <1$. For example, 
consider
\begin{eqnarray}
\rho_j = \frac{1}{2} (\mathds{1}+
\sum\limits_{\mu=x,y,z} \langle \sigma_\mu^j \rangle  \ 
\sigma_\mu^j) ,
\end{eqnarray}
where $\sigma_\mu^j$, $\mu=x,y,z$, are the
Pauli operators acting on the $j$-th spin,
\begin{equation}
\label{pauli3}
\sigma^j_\mu= \overbrace{\mathds{1} \otimes \mathds{1} \otimes \cdots \otimes
\underbrace{\sigma_\mu}_{j^{th}\ \mbox{factor}} \otimes \cdots
\otimes \mathds{1}}^{n\ \mbox{factors}} \; ,
\end{equation}
with
\begin{eqnarray}
\hspace*{-0.5cm}
\sigma_x=
\left (
\begin{array}{cc}
0 & 1 \\
1 & 0 \\
\end{array} \right ) , 
\sigma_y=
\left (
\begin{array}{cc}
0 & -i \\
i & 0 \\
\end{array} \right )  , 
\sigma_z=
\left (
\begin{array}{cc}
1 & 0 \\
0 & -1 \\
\end{array} \right ) ,
\end{eqnarray}
and $\langle
\sigma_\mu^j \rangle =\langle \Psi | \sigma_\mu^j | \Psi \rangle$ 
denotes the corresponding expectation value. The
vectors ${\bf X}_j = (\langle \sigma_x^j \rangle, \langle \sigma_y^j
\rangle, \langle \sigma_y^j \rangle )$, ${\bf X}_j \in {\mathbb R}^3$,
allow a geometric representation of each reduced state in
${\mathbb R}^3$, satisfying $0\le \|{\bf X}_j\| \le 1$. 
 Since ${\sf tr}_j (\rho^2_j) = \frac{1}{2} (1+\|{\bf X}_j\|^2)$,
 the state $\ket{\Psi}$ is entangled if $\|{\bf X}_j \| <1$
for at least one $j$, represented by a point {\em inside} the
corresponding local Bloch sphere. One may therefore consider
$\ket{\Psi}$ to be maximally
entangled if $\|{\bf X}_j \|= 0$ for all $j$. On the other hand, the state
$\ket{\Psi}$ is unentangled (i.e., a product state) if
$\|{\bf X}_j \| =1$ for all $j$, corresponding to points lying on the
surface of the Bloch sphere.

A natural geometric measure of multipartite entanglement
is obtained by defining the {\em purity of a state relative to a set
of observables} \cite{Barnum2003, Barnum2004}. If the set 
is chosen to be the set of {\em
all local observables}, i.e., corresponding to each of the
subsystems that compose the actual system, one recovers 
the standard notion of entanglement for multipartite systems. 
For example, if the system
consists of $n$ qubits, we obtain a measure of conventional
entanglement by calculating the purity relative to the set $\fh =
\{\sigma_x^1,\sigma_y^1, \sigma_z^1, \ldots,
\sigma_x^n,\sigma_y^n,\sigma_z^n \}$, 
\begin{equation}
P_\fh = \frac{1}{n} \sum\limits_{j=1}^{n}\sum\limits_{\mu=x,y,z}
\langle \sigma_\mu^j \rangle^2\:, \ \  0 \leq P_\fh \leq 1 \ .
\label{purityMeasure.eq}
\end{equation}
Since
$\fh$ is a semi-simple Lie algebra, its generalized unentangled states
are the generalized coherent states obtained by applying any 
group operation to a reference
state such as $\ket{\sf 0}=\ket{0} \otimes \cdots \otimes
\ket{0}$.
For the algebra $\fh$ of local observables, such
group operations are simply local rotations on each qubit. In other words,
the group orbit describing the generalized coherent states of $\fh$ comprises all
the product states of the form
$\ket{\Psi}=\ket{\psi_1}\otimes \cdots \otimes \ket{\psi_n}$, which
have maximum purity (i.e., $P_\fh=1$). Other states such as the 
Greenberger-Horne-Zeilinger state
$\ket{\Psi}=\ket{{\sf GHZ}_n}=\frac{1}{\sqrt{2}} ( \ket{0}
\otimes \cdots \otimes \ket{0} + \ket{1} \otimes
\cdots \otimes \ket{1})$ are (maximally) 
entangled relative to the set of local observables (i.e., $P_\fh=0$).

Different entanglement measures are obtained when a set $\fh$ different from the
local observables is chosen.  An obvious example, in particular, is
given by the set of all observables.  In this case, the purity takes
its maximum value independently of the pure quantum
state \cite{Barnum2003, Barnum2004}, expressing the fact that any state
is a generalized coherent state of the Lie algebra of all observables. 



\section{Vector Spaces over Complexified Finite Fields}

In order to address the intrinsic problems induced by the notion of
the continuum calculations of the previous section, one is led to
replace the infinite fields of CQC by discrete computable fields.
Accomplishing this while maintaining the essential elements of
addition and multiplication requires a brief excursion into the theory
of fields, and particularly the theory of finite fields. 

\subsection{Background} 

Abstract algebra deals with various kinds of algebraic structures,
such as groups, rings, and fields, each defined by a different system
of axioms. A field~$\mathbf{F}$ is an algebraic structure consisting
of a set of elements equipped with the operations of addition,
subtraction, multiplication, and division \cite{numtheory.ref}.  Fields
may contain an infinite or a finite number of elements. The rational
$\mathbb{Q}$, real $\mathbb{R}$, and complex numbers $\mathbb{C}$ are
examples of infinite fields, while the set $\mathbf{F}_3=\{ 0,1,2\}$
under the usual multiplication and modular addition is an example of a
finite field. Finite fields are also known as Galois fields
\cite{GT.ref}.

There are two distinguished elements in a field, the addition identity
$0$, and the multiplication identity $1$.  Given the 
field $\mathbf{F}$, the
closed operations of addition, ``$+$'', and multiplication,
``$\gmult$'', satisfy the following set of axioms:

\begin{enumerate}
\item 
$\mathbf{F}$ is an Abelian group under the addition operation $+$
(additive group) 

\item The multiplication operation $\gmult$ is associative and
  commutative. The field has a multiplicative identity and the
  property that every nonzero element has a multiplicative inverse

\item
Distributive laws: For all $a,b,c \in \mathbf{F}$
\begin{eqnarray}
a \gmult (b+c)=a \gmult b+ a
\gmult c \\ 
(b+c) \gmult a = b \gmult a+ c \gmult a  \ .
\end{eqnarray}
 
\end{enumerate} 
{}From now on, unless specified, we will omit the symbol $\gmult$
whenever we multiply two elements of a field.

Finite fields of $q$ elements, $\Fq=\{0,\ldots,q-1\}$, will play a
special role in this work.  A simple explicit example
is the following addition and
multiplication tables for $\mathbf{F}_3$:
\[\begin{array}{c|ccc}
+ & 0& 1 & 2\\[.1in] \hline
0 &  0 &  1  & 2 \\
1  &  1& 2 & 0 \\
2  & 2  & 0 & 1
\end{array}    \hspace*{2cm}
\begin{array}{c|ccc}
\gmult  & 0 & 1 & 2 \\[.1in] \hline
0 &  0 &  0  & 0 \\
1  &  0& 1 & 2 \\
2  & 0  & 2 & 1
\end{array} 
\]

\subsection{Cyclic Properties of Finite Fields}
Finite fields are classified by size.  The characteristic of a field is
the least positive integer $m$ such that $m=1+1+1+\cdots+1\equiv 0$,
and if no such $m$ exists we say that the field has characteristic
zero (which is the case for infinite fields such as $\mathbb{R}$). It turns out that if
the characteristic is non-zero it must be a prime $p$.  For every
prime $p$ and positive integer $r$ there is a finite field
$\Fpx{p^r}$ of cardinality $q=p^r$ and characteristic $p$
(Lagrange's theorem), which is unique up to field isomorphisms.  For
every $a \in \Fq$, $a \neq 0$, then $a^{q-1}=1$, implying the Frobenius
endomorphism (also a consequence of Fermat's little theorem)
$a^{q}=a$, which in turn permits us to write the
multiplicative inverse of any non-zero element in the field as
$a^{-1}=a^{q-2}$, since $a^{q-2}a=a^{q-1}=1$. 
Every subfield of the field $\Fq$, of
cardinality $q=p^r$, has $p^{r'}$ elements with some $r'$ dividing $r$,
and for a given $r'$ it is unique.  Notice that a fundamental difference
between finite and infinite fields is one of topology:  finite fields
induce a compact structure because of their modular arithmetic, 
permitting {\it wrapping around}, while
that is not the case for fields of characteristic zero. This feature may lead to
fundamental physical consequences.

\subsection{Complexified Finite Fields}

Consider the polynomial $x^2+1=0$ over a finite field $\Fp$. It is known that
this polynomial does not have solutions in the field precisely when the prime
$p$ is congruent to $3 \imod{4}$ (see, e.g., \cite{numtheory.ref}).

For such primes, it is therefore possible to construct an extended field
$\Fpp$ whose elements are of the form $\alpha=a+ib$ with $a \in \Fp$, $b \in
\Fp$, and $i$ the root of the polynomial $x^2+1=0$. Since the field elements
$a+ib$ behave like discrete versions of the complex numbers, we will refer to
fields $\Fp$ with prime $p$ congruent to $3 \imod{4}$ as \emph{complexifiable
  finite fields}, and $i$-extended fields $\Fpp$ with $p$ congruent
to $3 \imod{4}$ as \emph{complexified finite fields}.

In a complexified finite field $\Fpp$, the Frobenius automorphism that
maps $\alpha \in \Fpp$ to $\alpha^p \in \Fpp$ acts like complex
conjugation.  For example, in $\Fppx{3}$, we have $(2+i)^3 = 8 + 12i
-6 -i = 2 + 11i$ which, in the field, is equal to $2-i$ since $11
\equiv -1 \imod{3}$.

We define
the \emph{field  norm} $\fnorm{\cdot}$ as the map from $(a+ib) \in
\Fpp$ to $a^2+b^2 \in \Fp$,
\begin{equation}
 \fnorm{\alpha=a + i b} = a^2 + b^2 \ .
\label{fieldnorm.eq}
\end{equation}
We avoid the square root in the discrete field framework because,
unlike the continuous case, the square root does not always exist.

\subsection{Vector Spaces} 

In this section we want to build a theory of discrete vector spaces
that  approximates as closely as possible the
features of conventional quantum theory.  Such a structure would ideally
consist of the following:
(i) a vector space over the field of complex numbers, and (ii) an inner
product $\ip{\Phi}{\Psi}$ associating to each pair of vectors a complex
number, and satisfying the following properties:
\begin{itemize}
\item[A.] $\ip{\Phi}{\Psi}$ is the complex conjugate of $\ip{\Psi}{\Phi}$; 
\item[B.] $\ip{\Phi}{\Psi}$ is conjugate linear in its first argument and
linear in its second argument;
\item[C.] $\ip{\Psi}{\Psi}$ is always non-negative and is equal to~0 only
if $\ket{\Psi}$ is the zero vector. 
\end{itemize}
It turns out that a vector space defined over a finite field cannot have an
inner product satisfying the properties above. However, we will introduce an
Hermitian ``dot product'' satisfying some of those properties.

We are interested in the $n$-qubit vector space $\cal H$ of dimension $D=2^n$ defined
over the complexified field $\Fpp$.  Let $\ket{\Psi} =
(\alpha_0~\alpha_1~\ldots~\alpha_{D-1})^T$ and $\ket{\Phi} =
(\beta_0~\beta_1~\ldots~\beta_{D-1})^T$ represent vectors in $\cal H$, with
numbers $\alpha_i$ and $\beta_i$ drawn from $\Fpp$, and where $(\cdot )^T$ is the
transpose.

\begin{definition}[Hermitian dot product] 
  Given vectors $\ket{\Phi}$ and $\ket{\Psi} \in {\cal H}$, the Hermitian dot
  product of these vectors is:
\begin{equation}
\ip{\Phi}{\Psi} = \sum_{i=0}^{D-1} ~\beta_i^p~\alpha_i^{\;} \ .
\end{equation}
\end{definition} 
Two vectors $\ket{\Phi}$ and $\ket{\Psi} \in {\cal H}$ are said to be
orthogonal if $\ip{\Phi}{\Psi} =0$.  This product satisfies conditions A and 
B for inner products but violates
condition C since in every finite field there always exists a non-zero vector
$\ket{\Psi}$ such that $\ip{\Psi}{\Psi} = 0$.  The reason is that addition in
finite fields eventually ``wraps around'' (because of their cyclic or modular
structure), allowing the sum of non-zero elements to be zero.  The fraction
of non-zero vectors satisfying $\ip{\Psi}{\Psi} = 0$ decreases with the order
$p$. 

For any vector $\ket{\Psi} = (\alpha_0~\alpha_1~\ldots~\alpha_{D-1})^T$, the
Hermitian dot product $\ip{\Psi}{\Psi}$ is equal to $\sum_{i=0}^{D-1}~
\fnorm{\alpha_i}$, which is the sum of the field norms for the complex
coefficients.  For convenience, we now extend the field norm to
include vector arguments by defining 
\begin{equation}
\fnorm{\ket{\Psi}} = \ip{\Psi}{\Psi} =
\sum_{i=0}^{D-1}~\fnorm{\alpha_i} \ .
\label{fnormD.eq}
\end{equation}
The field norm of a vector can vanish for non-vanishing vectors.

For vectors $\ket{\Psi}$ such that $\ip{\Psi}{\Psi} = 
\sum_{i=0}^{D-1} \alpha_i^p \, \alpha_i^{\;}=
\sum_{i=0}^{D-1} \alpha_i^{p+1}=
\sum_{i=0}^{D-1} |\alpha_i|^2$ has a square root in the
field, one can define the following ``norm'':
\begin{equation}
\label{norm-eq}
||\Psi|| = \sqrt{\ip{\Psi}{\Psi}} \ ,
\end{equation}
which is valid only on a subspace of the field norm for finite fields.


\section{Irreducible Discrete $n$-qubit States: 
  Generalized Discrete Bloch  Sphere} 
\label{DQCnqubitBloch.sec}

\begin{figure}[htb]
\begin{center}
\hspace*{-0.5cm}\includegraphics[width=1.1\columnwidth]{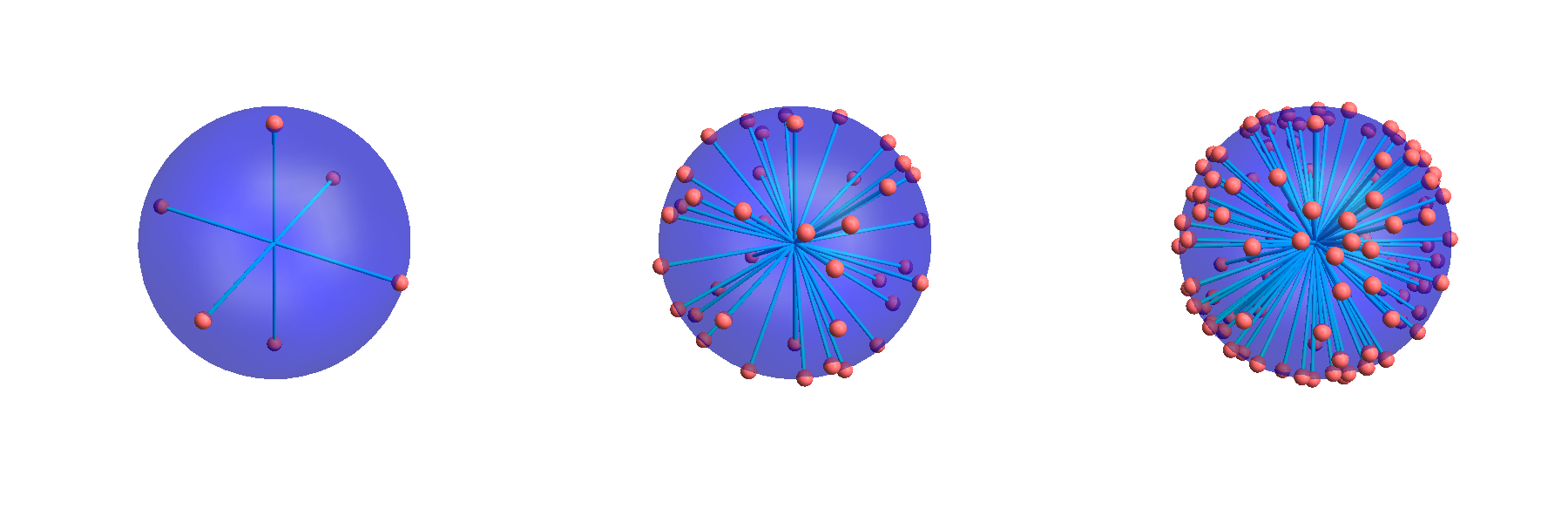}
\vspace{-0.2in}  
\end{center}
 \caption{Schematically normalized plots of the elements of the
   discrete Bloch sphere, the irreducible single-qubit
   (two-dimensional) state vectors  with unit norm over the field
   $\Fpp$.  We show the results for $p=3$, $7$, and $11$.
For example, in $\Fppx{3}$, there are 24 vectors of unit norm, but only
the 6 inequivalent classes appear in the plot.  The $p+1=4$
equivalent vectors in each class differ only by a complex discrete phase.}
 \label{bloch.fig}
\end{figure}

In the one-qubit state with coefficients in $\Fpp$, the discrete analog
of the Bloch sphere is constructed by exact analogy to the continuous
case: we first require that the coefficients of the single qubit
basis obey
\begin{equation}
\|\psi_{1}\|^{2}= | \alpha_{0}|^{2} +| \alpha_{1}|^{2} = 1 
\label{DQA2.eq}
\end{equation}
in the discrete field. In  \ref{appU.sec}, we show that there
are $p(p^2-1)$ such values.  Given this requirement, which is similar
in form to the conservation of probability, but not as useful due to
the lack of orderable probability values, we can immediately conclude
that the discrete analog of the Hopf fibration is again
\begin{eqnarray} 
 X & = & 2\, \mathrm{Re}\   \alpha_{0} \alpha_{1}^{*} = 2 x_0 x_1 + 2
 y_0 y_1 \nonumber \\
 Y & = & 2\, \mathrm{Im}\   \alpha_{0} \alpha_{1}^{*} = 2 x_1 y_0 - 2
 x_0 y_1  \label{DQA6.eq} \\ 
 Z & = &   | \alpha_{0}|^{2} - | \alpha_{1}|^{2} =  x_{0}^{\ 2} +
   y_{0}^{\ 2} - x_{1}^{\ 2} - y_{1}^{\ 2}  \ , \nonumber
\end{eqnarray}
but now with all computations in $(\mbox{mod}~p)$. At this point one simply
writes down all possible discrete values for the complex numbers $\{\alpha_{0}, 
\alpha_{1}\}$ satisfying Eq.~(\ref{DQA2.eq}) and enumerates those
that project to the same value of 
$\{X,Y,Z\}$.  This equivalence class is the discrete analog of the circle in
the complex plane that was eliminated in the continuous case.  
In \ref{appH.sec}, we show that
$p+1$ discrete values of $\{\alpha_{0}, \alpha_{1}\}$ with unit norm  map to
the same point under the Hopf map \Eq{DQA6.eq}; we may think of these as
discrete circles or projective lines of equivalent, physically
indistinguishable, complex phase. 
The surviving $p(p-1)$ values of $\{\alpha_{0}, \alpha_{1}\}$ correspond to
irreducible physical states of the discrete single qubit system. 
Thus, for example, choosing the underlying field to be $\Fppx{3}$,
there are exactly 6 single-qubit state vectors to populate the Bloch
sphere; the four equivalent phase-multiples mapping to each of the six
points on the $\Fppx{3}$ Bloch sphere are collapsed and regarded as physically
indistinguishable.   In Figure
\ref{bloch.fig}, we plot the irreducible states on the Bloch sphere for $p=3$,
$7$, and $11$.  Note that the Cartesian lengths of the real vectors
corresponding to the points on the Bloch sphere vary considerably due
to the nature of discrete fields; we have artificially normalized them to
a ``continuous world'' unit radius sphere for conceptual clarity.

\subsection{Counting states on the $n$-qubit Bloch sphere}

We have the unique opportunity in the finite-field approach to
quantum computing to precisely identify and enumerate the physical
states.  In the conventional theory, as we have seen, we employ a
generalized Hopf fibration on the normalized states to project out
a circle of phase-equivalent states, yielding the generalized Bloch
sphere.

In the introduction to this section, we sketched the counting of the
irreducible single-qubit discrete states.
To count the number of inequivalent discrete states for the general
$n$-qubit case with coefficients in $\Fpp$, we first must find the set
of unit-norm states, and then determine the equivalence classes of
unit-norm states under discrete 
phase transformations; we can then enumerate the list of states on the
discrete generalized Bloch sphere.  By executing computer searches of these
spaces, we discovered an hypothesis for a closed-form solution for the
counting of the states, and were then able to find a rigorous inductive proof
of the enumeration, which is presented in the Appendix.

This process of describing the discrete $n$-qubit irreducible states
can again be understood geometrically by following the discrete analog
of the Hopf fibration.  First, we construct the discrete version of the quadratic
unit-length form that automatically annihilates the distinction
among states differing only by a discrete phase,
\begin{eqnarray}
\hspace*{-0.5cm}
\Vec{X} = ( |\alpha_{i}|^{2},..., \sqrt{2}\   \mathrm {Re}\ \alpha_{i}
\alpha_{j}^{*}, \ldots, \sqrt{2}\  \mathrm{Im}\   \alpha_{i}
\alpha_{j}^{*},\ldots )\ ,
\end{eqnarray}
where
\begin{eqnarray}
\Vec{X}\cdot\Vec{X} = \left( \sum_{i=0}^{D-1} \left |
    \alpha_{i}\right | ^{2}\right) ^{2} = 1 \ .
\end{eqnarray}
{}From  \ref{appH.sec}, we know that $p+1$ elements of this
discrete $\Sphere{2\times2^{n} -1}$ structure map to the {\it same
  point\/} in $\Vec{X}$.  Each set of $(p+1)$ redundant points is,
geometrically speaking, the {\it discrete Hopf fibration circle\/}
living above each {\it irreducible\/} point of the $n$-qubit state
description.  These $p+1$ points are interpretable as the $p$ finite
points plus the single point at infinity of the projective discrete
line (see, e.g., \cite{Arnold}).

The next part of this argument is the determination of the unit-norm states,
effectively the space of allowed discrete partitions of unity; we cannot
exactly call these ``probability-conserving'' sectors of the state
coefficients since we do not have a well defined notion of probability, but
we do have a well-defined notion of partition of unity.  The tally of
unit-norm states is $p^{2^{n}-1}(p^{2^{n}} -1)$ (see 
\ref{appU.sec}) compared 
to the total number $p^{2\times2^{n}}$ of possible complex integer state
vectors that could be chosen.  This unit-norm state structure is the
discrete analog of $\Sphere{2\times2^{n} -1}$.

Finally, we repeat the last step of the $n$-qubit continuous Hopf
fibration process for discrete $n$-qubit states, eliminating the
discrete set of $p+1$ equivalent
points that map to the same point $\Vec{X}$ on the generalized
$n$-qubit Bloch sphere.  Dividing the tally
$p^{2^{n}-1}(p^{2^{n}} -1)$ of unit norm states by the $p+1$ elements
of each phase-equivalent discrete circle, we find
\[\frac{ p^{2^{n}-1}(p^{2^{n}} - 1)}{p+1} =
p^{2^{n}-1} (p-1)\,\prod_{k=1}^{n-1} ( p^{2^{k}}+1) \] as the total
count of unique irreducible states in a discrete $n$-qubit
configuration (see  \ref{appI.sec}).  The resulting object is
precisely the discrete version of $\CP{D-1}$, which we might call a
{\it discrete complex projective space\/} or $\DCP{D-1}$, where  
$D=2^{n}$ as usual.

\section{Geometry of Entangled States}

\begin{figure}[htb]
\begin{center}
\includegraphics[width=1.0\columnwidth]{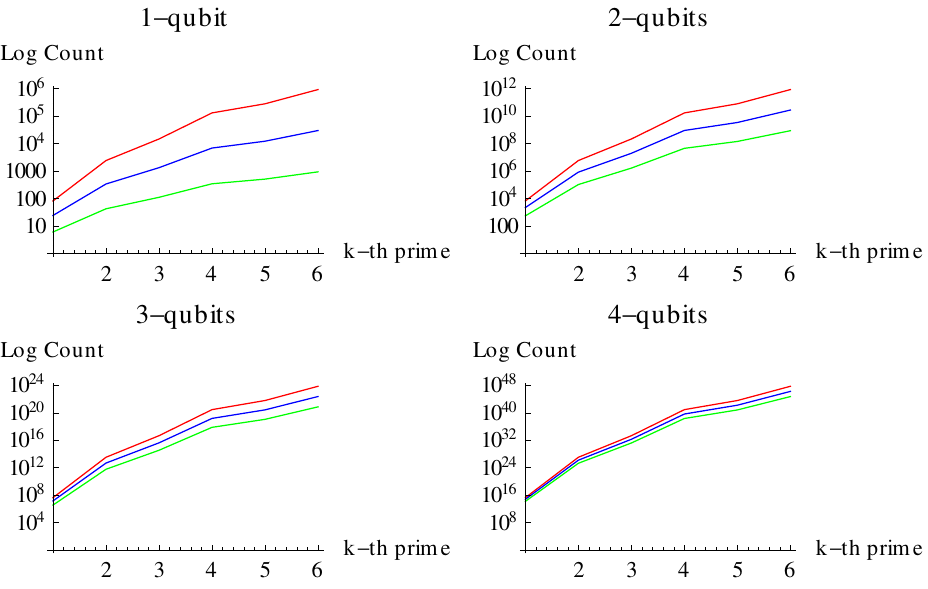}
\end{center}
 \caption{Logarithmic plot of the number of discrete unnormalized
states (top, in red), vs the number of normalized discrete states
(middle, in blue), vs the irreducible states (bottom, in green)
 for the first 6 $\Fpp$-compatible primes,  (3, 7,
11, 19, 23, 31), for the number of qubits 1, 2, 3 and 4. }
 \label{statePlot3Unit.fig}
\end{figure}

Without regard to uniqueness, an $n$-qubit state with discrete
complex coefficients in $\Fpp$ will have the total possible space of
coefficients with dimension $p^{2\times 2^{n}}$ (including the null state).  
Imposing the condition of a length-one norm 
in $\Fp$, this number is reduced to $p^{2^{n}-1}(p^{2^{n}} - 1)$.  The ratio
of all the states to the unit-norm states is asymptotically $p$:
\begin{eqnarray}
\frac{p^{2^n+1}}{p^{2^n}-1} \rightarrow p \ , 
\end{eqnarray}
so there are roughly $p$ sets of coefficients, for any number of
qubits $n$, that are discarded for each retained unit-length state
vector. A factor of $p+1$ more states are discarded in forming the
discrete Bloch sphere of irreducible states.
Selected plots of the full space compared to both the
unit-norm space and the irreducible space for a selection of
complexified finite fields are 
shown in Figure \ref{statePlot3Unit.fig} for 1, 2, 3, and 4 qubits.



\subsection{Unentangled vs Entangled Discrete States}

For a given $p$ and the corresponding complexified field $\Fpp$, the
$n$-qubit discrete quantum states with coefficients in $\Fpp$ can be
classified by their degree of entanglement to a level of precision
that is unavailable in the continuous theory.  We look first at the
unentangled $n$-qubit states, which are direct product states of the
form 
\begin{equation} 
\ket{\Psi} = \ket{\psi_1}\otimes \cdots \otimes \ket{\psi_j} \otimes \cdots 
\otimes \ket{\psi_n}\ .
\label{dqcNqUntg.eq}
\end{equation}
Without regard to normalization, there are $(p^{4})^{n}$ possible
unentangled states out of the total of  $p^{2\times 2^{n}}$ states noted
above.  When we normalize the individual product states to unit norm,
the norm of the entire $n$-qubit state becomes the product of those
unit norms, and is automatically normalized to one.  We have already
seen that each single-qubit normalized state in the tensor product
\Eq{dqcNqUntg.eq} has precisely $p(p-1)$ irreducible components.

\subsection{Completely Unentangled States and the Discrete Bloch
  Sphere}  In effect, the irreducible states for unentangled $n$-qubit
configurations reduce to a single Bloch sphere for each one-qubit
component $\ket{\psi_{j}}$, and thus the whole set of states is defined
by an $n$-tuple of discrete Bloch sphere coordinates.  Since each
Bloch sphere in $\Fpp$ has $p(p-1)$ distinct irreducible components,
we have
\[ \mbox{\bf Count of Unentangled States} = p^{n} (p-1)^{n} \ .\]

We know that the total number of irreducible states (points in the
generalized $\DCP{2^{n}-1}$ Bloch sphere) for an $n$-qubit state is
$p^{2^{n}-1}(p^{2^{n}} - 1)/(p+1)$, and so the number of states
containing some measure of entanglement is
\begin{eqnarray}
\mbox{\bf Count of Entangled States} =\frac{p^{2^{n}-1}(p^{2^{n}} - 1)}{p+1} -  p^{n} (p-1)^{n} \ . \nonumber
\end{eqnarray}
Therefore a very small fraction of the unit norm states are 
unentangled.

\subsection{Partial entanglement}  A partially entangled state can be
constructed by taking individual component states to enter as direct
products, starting by picking $n$ distinct single qubits to be
unentangled ($n=2$ exhausts its freedom with one pick).  Then we can
choose $n(n-1)/2$ pairs of distinct qubits as product states ($n=3$
exhausts its freedom with one pick), and so on.  Then, starting with
$n=3$, we can pick fully entangled 2-qubit subspaces as product
spaces, combining them with single-qubit product components ($n=3$ has
no single-qubit freedom left after picking any of its three 2-qubit
subspaces), and so on.  Precise measures of entanglement such as that
given in \Eq{purityMeasure.eq} can then be applied just as in the
continuous case.

\subsection{Maximal entanglement}
Numerical and analytic calculations of the entanglement measure \Eq{purityMeasure.eq},
taken $(\mbox{mod}~p)$, extend to the best of our knowledge to the discrete
case, so that the unentangled states constructed above have $P_{\fh}=1$.
This leads us to study one final aspect of the discrete $n$-qubit states,
namely the {\it maximally\/} entangled states with $P_\fh=0$.

Computing some examples for various $n$ and small values of $p$, one can
verify explicitly that unit-norm  unentangled states for $n=2$,
$p = \{3, 7, 11, 19,\ldots\}$ occur with frequency
\[ (p + 1) p^{2} (p - 1)^2 = \{144, 14112, 145200, 2339280, \ldots \} \
, \]
and for general $n$, $(p+1) p^{n} (p - 1)^n$.

The irreducible state counts are reduced by $(p+1)$, giving
\[  p^{2} (p - 1)^2 = \{36, 1764, 12100, 116964, \ldots \} \
, \]
and in general for $n$-qubits, $ p^{n} (p - 1)^n$ instances
of pure states with $P_{\fh}=1$.

Repeating the computation to discover the frequency of maximally
entangled (purity  $P_\fh=0$ states), we find  $ p^{n+1} (p-1)
(p+1)^n$  maximally entangled states, with example frequencies
for two qubits of
\[ p^{3} (p-1) (p+1)^2 = \{864, 131712, 1916640, 49384800,\ldots\} \
. \]

The irreducible state counts for maximal entanglement are reduced by
$(p+1)$, giving for $n=2$
\[  p^{3} (p^2 - 1) = \{216, 16464, 159720, 2469240, \ldots \} \
, \]
and in general for $n$-qubits, $p^{n+1} (p-1) (p+1)^{n-1}$ instances
of pure states with $P_{\fh}=0$.

Therefore, the ratio of maximally entangled to unentangled states is
\[ \mbox{\bf Max Entangled/Unentangled} = p \left (\frac{p+1}{p-1} 
\right )^{n-1} \ .\]

\section{Summary}  Given a discrete basis for the complex
coefficients of an $n$-qubit quantum state, DQC permits us in
principle to explicitly determine the relative frequencies of phases
and to determine exactly the generalized Bloch sphere coordinates of
the irreducible states.  The size of the set of states that must be
taken as equivalent to get irreducibility is the size of the
``circle'' or phase group, and this is $p+1$ for any $p$ and for any
$n$ (related to the size of the finite projective line, see \cite{Arnold}).  
Exploring the discrete manifestation of the
purity measure \Eq{purityMeasure.eq}, our DQC approach can determine
not only the size of the irreducible space of states, but also the
relative sizes of the unentangled and entangled states for $n$
discrete qubits.

\appendix

\section{Proofs}

In this appendix, we prove the state-counting formulas for discrete
$n$-qubit states labeled by a prime $p$ satisfying $p= 4 \ell + 3$ for
integer $\ell\ge 0$.  We show that: (1) the number of points on a
discrete complex unit circle, (the discrete complex phase equivalence)
is $p+1$; (2) the number of unit-length $D$-dimensional vectors with
coefficients in $\Fpp$ is $ p^{D-1}(p^D - (-1)^D)$; for $n$-qubit
states, $D=2^{n}$ and the result becomes $p^{2^{n}-1}\left(
  p^{2^{n}}-1\right)$; and (3) the number of irreducible $n$-qubit
states is $p^{2^{n}-1} (p-1)\,\prod_{k=1}^{n-1} (p^{2^{k}}+1)$.

 We will carry out an inductive proof  
starting with an hypothesis for the number of zero-norm
$D$-dimensional vectors suggested by computing representative
examples.  We will accomplish this by exploring the properties of
finite fields using the one-dimensional and $n$-dimen\-sional field
norms over $\Fpp$ defined in Eqs.~(\ref{fieldnorm.eq}) and
(\ref{fnormD.eq}), that is,
\[\fnorm{\alpha =a+ib}=a^2+b^2\] and
\[\fnorm{\alpha_0,\ldots,\alpha_{D-1}} \equiv \fnorm{D}=
\sum_{k=0}^{D-1}~\fnorm{\alpha_k} \ ,\] 
where we will specify real discrete values using roman letters such as
$(a,b,c)$ and complex discrete values using  greek letters such as
$\alpha$, which stands for $\alpha = a + i b$.
We carry out the calculations for arbitrary $D$, and then 
specialize at the end to the even-$D$, $n$-qubit case
$D = 2^{n}$.     For additional general background, see, e.g.,
Chapter VI in \cite{numtheory.ref} and Section 18.4 in \cite{GT.ref}.

\subsection{Counting of the quadratic map}
\label{appH.sec}
{\bf Proposition 1:}  The discrete analog of phase-equivalence under $z
\rightarrow e^{i \phi} z$ is a set of $(p+1)$ discrete
points $\alpha \in \Fpp$ that map to unity in $\Fp$
under the action of $\fnorm{\alpha}$.\\[.05in]

{\bf Method:} To prove Proposition 1, we start by defining a special
case of the field norm $\fnorm{.}$, namely the real quadratic map
$Q(e)=e^2$ taking an arbitrary element $e \in \Fp$ to its square in
the field. We exploit the fact that the image of $Q(e)$ has $(p+1)/2$
unique elements in $\Fp$, including the zero element; the map
$Q^{*}(e)$ excluding the zero element produces $(p-1)/2$ elements (the
quadratic residues); the $(p-1)/2$ remaining elements of $\Fp$ (the
quadratic non-residues) are analogous to negative numbers, having no
square roots in the field $\Fp$.

{\bf Proof:}  We let $A$ be the image
of the map $Q(e)$ in $\Fp$, and note that the set $A_{c}$ resulting
from displacing an element $x=b^2$ of $A$ to $c-x= c - b^2$ with $c
\in \Fp$ also has $(p+1)/2$ unique elements because the result is
simply a cyclic shift of element labels.
We now observe that for any non-zero $c \in \Fp$, the join of the 
two sets $A$ and $A_{c}$ has size $p+1$, which is greater than the
size $p$ of $\Fp$, and so there must be at least one common element
such that 
\[ a^2 = c- b^2  \ . \]
Thus some element $c \in \Fp$ is the field norm of some element
$\alpha = a + i b \in \Fpp$, 
\[ \fnorm{\alpha} = a^2 + b^2 = c \ .\]
Since we required $c$ to be non-zero, and $\fnorm{\alpha=a + i b} = 0$ only
for $a=b=0$, the corresponding element $\alpha \in \Fpp$ must be non-vanishing.

This shows that for any non-zero  element $c \in \Fp$, there exists a
non-vanishing element $\alpha \in \Fpp$ with $\fnorm{\alpha} = c$, 
and thus we find
that the map $\fnorm{\alpha} : \alpha \in \Fpp \rightarrow c\in \Fp$ is
onto; in addition, since we could displace $c$ to any element of
$\Fp$, each non-zero element in the range $\Fp$ of the map
$\fnorm{\alpha}$ must correspond to the same number of non-zero
domain elements $\alpha \in \Fpp$.  Restoring the zero-element case,
we see also that  no elements of the full set of  $\Fp$ are missed in
the range of $\fnorm{\alpha}$.  

\begin{figure}[t]
\centerline{
\psfig{width=3.6in,figure=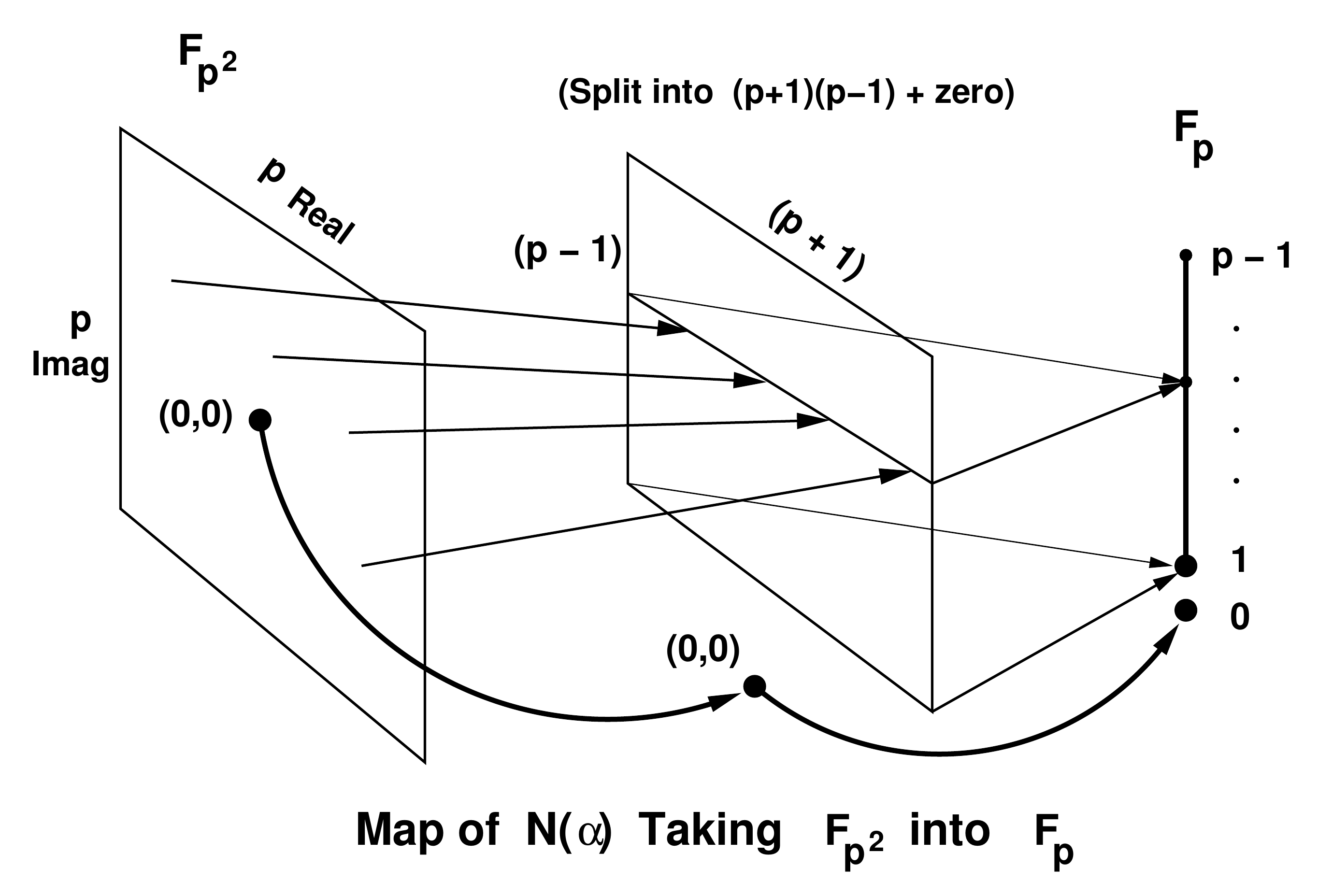}}
\caption[]{Sketch of the map from $\Fpp$ to $\Fp$ using
  $\fnorm{\alpha}$,
showing the decomposition of $\Fpp$ into the zero element $(0,0)$ and
the $p^2-1=(p+1)(p-1)$ non-zero elements that map onto the $(p-1)$
non-zero elements of $\Fp$ with multiplicity $(p+1)$.
}
\label{proof1.fig}
\end{figure}

We can now compute the size of the equivalence class of complex
unit-modulus phases corresponding to the Hopf fibration circle.  Since
$\Fpp$ has $p^2-1$ non-zero values, and the map $\fnorm{\alpha}$
distributes these equally across the domain of $p-1$ non-zero elements
$c \in \Fp$, there are $(p^2-1)/(p-1) = p+1$ (non-zero) domain
elements in $\Fpp$ for each (non-zero) image element in $\Fp$.  We
illustrate this graphically in Figure \ref{proof1.fig}.  Thus the Hopf
circle always has size $p+1$, corresponding essentially to a discrete
projective line, and that is the size of each equivalence class of the
map $\fnorm{\alpha}$ for non-vanishing $\alpha$, including in
particular the map to the unit norm value $ c=1 \in \Fp$.

\subsection{Counting of Unit-Norm states}
\label{appU.sec}
{\bf Proposition 2:} The number of unit-norm 
states described by a $D$-dimensional vector
$(\alpha_0,\ldots,\alpha_{D-1})$ with coefficients $\alpha_{i} 
\in \Fpp$ is $\omega(D,p) = p^{D-1} (p^D - (-1)^D)$.\\[.05in]

{\bf Method:} We generalize Proposition 2 to also provide the count of the
zero-norm states $\zeta(D,p) = p^{D-1}(p^D + (-1)^D (p-1))$ and prove both
formulas simultaneously by induction on $D$. 

{\bf Proof:} The field-norm map
$\fnorm{\alpha_0,\ldots,\alpha_{D-1}}=\fnorm{D}: (\Fpp)^{D} \rightarrow \Fp$
takes the domain of $D$-dimensional vectors, with total number of possible
cases $(p^{2})^{D} = p^{2 D}$, to an image of discrete size $p$ in $\Fp$,
which we can think of either as a zero-origin set $\{0,1,\ldots,p-1\}$ or as
a zero-centered set $\{(-(p-1)/2,\ldots,-1,0,1,\ldots,(p-1)/2\}$.  The latter
is useful for considering pairings of numbers that sum to zero in the field
$\Fp$.

{\it Zero-norm case.\/}   We begin with our experimentally
generated hypothesis for the {\it number of zero-norm vectors\/} with
no restriction on the parity of $D$, allowing $D+1$ to be odd as well
as even:
\begin{equation}
 \zeta(D,p) =  p^{D-1}(p^D + (-1)^D (p-1)) \ .
\label{zeronorm.eq}
\end{equation}
This is the proposed number of values of  $(\alpha_0,\ldots,\alpha_{D-1}) \in
(\Fpp)^{D}$ for which $\fnorm{D} = 0$.

{\it Unit-norm case.\/} 
Next, we observe that, since there are $p^2$ elements $\alpha \in \Fpp$,
we must have $(p^{2})^{D}=p^{2D}$ possible values of a $D$-dimensional
vector $(\alpha_0,\ldots,\alpha_{D-1})$.  There are  $p^2-1$ non-zero
values of  $\alpha \in \Fpp$, and we showed in Proposition 1 that
$\fnorm{\alpha}$ maps exactly $p+1$ values in that set to each
of the $p-1$ non-zero values in $\Fp$.
Therefore, we can propose that the {\it unit-norm case\/} has 
a count of domain elements that is $1/(p-1)$ of the total number of
non-zero-norm cases.   The proposed number of unit-norm cases
following from the hypothesis Eq.~(\ref{zeronorm.eq}) would then be
\begin{eqnarray}
 \omega(D,p)& =& \frac{ p^{2 D} - \zeta(D,p)}{p-1} \ .
\label{unitnorm.eq}
\end{eqnarray}

\paragraph*{Proof by Induction on $D$.}
Since, by Eq.~(\ref{unitnorm.eq}), the proposed unit-norm counting formula
$\omega(D,p)$ for a given $D$ follows immediately from the proposed zero-norm
counting formula $\zeta(D,p)$ for the same $D$, it is sufficient to perform
our inductive proof on the zero-norm counting formula implicitly using the
statement for the one-norm counting formula.  We thus assume that we are
given $\zeta(D,p)$, and proceed to examine the relation between the vanishing
domains of $\fnorm{D}$ and $\fnorm{D+1}$,
which can be written for generic $\alpha = \alpha_{D}$ as
\begin{equation}
\fnorm{D+1} = \fnorm{D} + \fnorm{\alpha} \ .
\label{dplus.eq}
\end{equation}
The  counting of elements in the domain of the  $\fnorm{D+1}$ map
whose image in $\Fp$ is zero consists of two parts:
\begin{itemize}
\item {\bf Simple zeroes.} If $\alpha = 0$, the only possible zeroes of
  $\fnorm{D+1}$ are the zeroes of   $\fnorm{D}$, counted by one
  instance of $\zeta(D,p)$.
\item {\bf Compound zeroes.} If  $\alpha \neq 0$, then
  $\fnorm{\alpha}= c$ for non-zero $c \in \Fp$.  As we noted,
the values of $c$ can be written as $(p-1)/2$ pairs of matched
positive and negative numbers that sum pairwise to zero in the field
$\Fp$.  However, we know that  $\fnorm{D}$ maps its domain to 
{\it each\/} value of non-zero $c \in \Fp$ exactly $p+1$ times.
Assuming that $\zeta(D,p)$ is true, we may use the resulting
hypothesis for the formula of
Eq.~(\ref{unitnorm.eq}) expressing $\omega(D,p)$,
the unit-norm counting hypothesis, directly in terms of $\zeta(D,p)$.
The compound zero counts then follow from using
$\omega(D,p)$ as the number of times that the {\it negated\/} value, that is
$-c$, is encountered to match each non-zero value of  $\fnorm{\alpha}=
c$.  Therefore we find that $p^2-1$ instances of the count
$\omega(D,p)$ would contribute to the final hypothesized tally of zeroes
of  $\fnorm{D+1}$. 
\end{itemize}

The inductive proof of Eq.~(\ref{zeronorm.eq}) then
proceeds by verifying the validity of the base case
\[\zeta(1,p)=1 \]
combined with the following verification of the
counting of the zeroes $\zeta(D+1,p)$ of $\fnorm{D+1}$ in terms of
$\zeta(D,p)$. 
\begin{eqnarray}
 \zeta(D+1,p)& =&  \zeta(D,p) + (p^2-1)\, \omega(D,p) \nonumber \\
& \stackrel{\mathrm{(\ref{unitnorm.eq})}}{=}&  \zeta(D,p)+ (p^2-1)\,\frac{ p^{2 D} - \zeta(D,p)}{p-1} \nonumber
\\
 & =&    p^{2 D+1} + p^{2 D} -  p\, \zeta(D,p) \nonumber \\
 & \stackrel{\mathrm{(\ref{zeronorm.eq})}}{=}  & p^{D}(p^{D+1} + (-1)^{D+1} (p-1))\ .
\label{fnormz.eq}
\end{eqnarray}
The result follows from observing that this
is the required form of Eq.~(\ref{zeronorm.eq}) for
$D\rightarrow D+1$.  Since   Eq.~(\ref{zeronorm.eq}) is the zero-norm
count for all $(D,p)$, a corollary is that  Eq.~(\ref{unitnorm.eq}) is
the count of unit-norm discrete states for all $(D,p)$.

\subsection{$n$-qubit formulas}
\label{appI.sec}
  Moving to the case of interest where
$D=2^n$ is the (even) state-vector length for an $n$-qubit state, we
have proven that the number of unit-norm states of an $n$-qubit vector
$\ket{\Psi}$ is
\[ \omega(2^{n},p) = p^{2^{n}-1}\left( p^{2^{n}}-1\right) \ .\]
 Since the multiplicity of points  $\alpha \in \Fpp$ mapping to
the same point, in particular the unit value,
 under the action of $\fnorm{\alpha}$ is $p+1$,
the number of irreducible discrete $n$-qubit states on the generalized
discrete Bloch sphere is simply the quotient
\begin{eqnarray*}\hspace*{-1cm}
\mbox{\rm \bf Irreducible $n$-qubit states} =
   \frac{p^{2^{n}-1}\left( p^{2^{n}}-1\right)}{p+1} 
=  p^{2^{n}-1} (p-1)\,\prod_{k=1}^{n-1} ( p^{2^{k}}+1) \ .
\end{eqnarray*}

\end{document}